 \documentclass[aps,preprint,groupedaddress,showpacs]{revtex4}
\usepackage{graphicx}

\begin{document}
\newcommand{\pH}{p-H$_2$\ }
\bibliographystyle{apsrev}


\title{Variational Monte Carlo study of the ground state properties and vacancy formation energy of  solid para-H$_2$ using 
a shadow wave function}



\author{Francesco Operetto}

\affiliation{Dipartimento di Fisica, Universit\`a di Trento, via Sommarive, 14
I-38050 Povo, Trento, Italy
}

\author{Francesco Pederiva}
\affiliation{Dipartimento di Fisica, Universit\`a di Trento, via Sommarive, 14
I-38050 Povo, Trento, Italy
}
\affiliation{INFM DEMOCRITOS National Simulation Center, Trieste, Italy}


\date{\today}

\begin{abstract}
A Shadow Wave Function (SWF) is employed along with Variational
Monte Carlo techniques to describe the ground state properties of
solid molecular para-hydrogen. The study has been extended to densities below 
the equilibrium value, to obtain a parameterization of the 
SWF useful for the description of inhomogeneous phases. 
We also present an estimate of the vacancy formation energy as a function of the 
density, and discuss the importance of relaxation effects near the vacant site.

\end{abstract}

\maketitle

\section{Introduction}
The properties of molecular para-hydrogen (\pH) have been the subject of intense 
experimental and theoretical investigation over the years. One main research
direction is related to the existence of a metallic phase at very large
pressures\cite{Hem90,vSt88,Sil80,Sil81,Sil90,Run92,Sur97}.  On the other hand, 
the Bosonic nature and the very small mass
of \pH would naturally lead to the search  for the existence of 
a superfluid phase. The strong H$_2$--H$_2$ interaction, however, 
causes the zero--pressure ground state of \pH to be a crystal at T=0.   
The only possibility of finding \pH in a fluid phase at temperatures lower than
6K\cite{Gin72}, where a superfluid transition is believed to occur,  
comes from  geometry--constrained or strongly inhomogeneous systems. 

It has been shown by Sinzingdre et al.\cite{Sin91} using Path--Integral Monte Carlo methods 
that small \pH clusters with a number of molecules N$\leq$13
preserve a liquid character, and show the signatures of superfluidity.
Recently Levi and Mazzarello\cite{Lev01} proposed that the existence of such liquid 
clusters might depend on the difficulty for the system to go over the nucleation
energy barrier in a time comparable to the lifetime of the clusters themselves.
The determination of such energy barrier depends on the knowledge of a
number of physical parameters, but in particular on the solid--liquid interface
energy.
It was recently shown that superfluid \pH could also be found in two 
dimensional layers deposited  on substrates such as graphite. 
Recent PIMC calculations showed that at full coverage the 
density of a monolayer is too large for allowing the system to have
a stable liquid phase. However, the presence of alkali impurities, that
exert a very weak attraction on the hydrogen molecules, can 
stabilize a disordered phase reducing the strong \pH-\pH
interaction\cite{Gor97,Nho02}. The same effect can be studied reducing the coverage and
looking at two--dimensional \pH  clusters at the surface. If the number of atoms
in a single cluster is small enough (N $\le$ 30) the stable configuration
is a puddle of  liquid\cite{Gor02}. 
The occurrence of surface melting at the surface of a thick (5 to 7 monolayers)
layer of \pH was also studied by PIMC techniques for temperatures $>$ 3K\cite{Wag96}. 

Recent experiments performed by Grebenev et al.\cite{Gre00} on a OCS--\pH immersed in 
$^4$He and mixed $^4$He--$^3$He clusters clearly show that the hydrogen
coating the molecule undergoes a transition at a temperature between 0.38K and 0.15K.
This transition can be inferred by the change in the measured rotational
spectrum, and therefore of the momentum of inertia, when the OCS--\pH complex is included in a  
pure $^4$He cluster and when also $^3$He is present. Recent simulation work confirms the
occurrence of this transition\cite{Kwo02}. 

From the theoretical point of view the simulation approach is the most effective in studying
inhomogeneous phases. In this paper we extend to the study of \pH a technique that revealed extremely powerful
in the description of inhomogeneous phases of $^4$He with large numbers of particles: 
Variational Monte Carlo calculations based on Shadow Wave functions (SWF)\cite{Vit88,Rea88}.
The main property of SWF is the capability of describing the crystalline and the disordered
phase of a quantum system with the same translationally invariant form. 
It was therefore possible to study coexisting 
liquid--solid\cite{Ped94} and liquid--vapor\cite{Gal98} systems in $^4$He as well as
finite systems like $^4$He clusters with and without impurities\cite{Gal01}. 
This method, although not exact, gives   
quantitatively significant results,  as for instance a realistic estimate of the
interface energy of solid $^4$He\cite{Ped94}, in situations where other techniques, like Path Integral MC
or Diffusion MC, are
harder to apply. Some early calculations using SWF for small \pH clusters (N$\leq$7) were
performed by Rama Krishna and Whaley\cite{RaK91}. However, their form of the wave function used 
cannot be immediately extended to bulk systems.
In this paper we introduce a SWF optimized and parametrized in a form directly
usable  in  calculations where the parameters become functions of the {\it local} density
of the system\cite{Ped94}. 
With this wave function, we carried out a systematic study of the ground state properties
of the  bulk solid  at densities up to 1.6 times the equilibrium density, and also 
to metastable bulk states below the equilibrium, beyond the limit of stability of the ordered phase.
We also present results for an estimate of the vacancy formation energy in \pH as a function
of the crystal density.

\section{Methods}
\subsection{Hamiltonian}
Molecular para-hydrogen is here described as a collection of $N$ point-like particles
interacting via a two-body potential:
\begin{equation}
\hat H = -\frac{\hbar}{2m}\sum_{i=1}^{N} + \sum_{i<j}v(r_{ij}).
\end{equation}
The model potential we chose is the Silvera-Goldman (SG)\cite{Sil78} potential, which takes the form:
\begin{equation}
v(r)=A[v_{rep}+(v_{att}+v_{dd})f_c(r)],
\end{equation}
where, using $\tilde r = r/r_m$.
\begin{eqnarray}
v_{rep}(r)&=&\exp(a - b \tilde r -\gamma \tilde r^2) \nonumber\\ 
v_{att}(r)&=& -\left(\frac{C_6}{\tilde r^6}+\frac{C_8}{\tilde r^8}+\frac{C_{10}}{\tilde r^{10}}\right) \nonumber\\
v_{dd}(r)&=& \frac{C_9}{\tilde r^9}\\
f_c(r)&=&\left\{ 
\begin{array}{l}
\exp\left[-\left(\frac{D}{\tilde r}-1\right)^2\right]\;\;{\rm if }\;\; r<D\;r_m\\ \\
1.0 \;\;{\rm if } \;\;r> D\; r_m
\end{array}
\right.. \nonumber
\end{eqnarray}
The values of the parameters are given in table I. 
\begin{table}
\begin{center}
\begin{tabular}{cr}
\toprule
$r_m$ (\AA)&3.41\\
$D$&1.28\\
$a$&10.923340\\
$b$&10.098343\\
$\gamma$&0.4122340\\
$C_6$&1.69550147\\
$C_8$&0.71379389\\
$C_9$&0.07468938\\
$C_{10}$&0.38990868\\
$A$ ($K$)&31.5763295\\
\botrule
\end{tabular}
\end{center}
\caption{Parameters of the SG potential\cite{Sil78}. All parameters are non--dimensional,
except for $r_m$, which is given in \AA, and $A$, which is given in K}  
\end{table}  

The $C_9/\tilde r^9$ term is introduced as 
an effective many--body force, assuming that the leading term is a triple-dipole
contribution, which is usually introduced as a three-body Axilrod-Teller force. However, 
this parameterization is  effective for studying isotropic properties of the system,
and the replacement of the three-body term with a pair term is a second order
effect. Variational Monte Carlo calculations by Norman et al.\cite{Nor84}, using a Bijl-Jastrow-Nosanow
\cite{Nos64}
trial wave function, show that within this model the GS potential is in qualitative
agreement with experiments in a wide range of densities;  other model interactions
(like the Schaefer-Watts potential\cite{Sch82} or the Buck et al. potential\cite{Buc83})
give better values around the equilibrium density, but become worse at higher pressure.

\subsection{Shadow Wave Function}

We use as trial solution for the N-molecules Hamiltonian at T=0K 
a Shadow Wave Function\cite{Vit88,Rea88}, 
having the following general form:
\begin{equation}
\Psi(R)=\int\;K(R,S)\phi(S)\;dS,
\end{equation}
where $R=\{{\bf r}_1\dots{\bf r}_N\}$ and $S=\{{\bf s}_1\dots{\bf s}_N\}$ are 
the molecular and auxiliary (``shadow'') degrees of freedom respectively.
The kernel $K$ is written as the product of a Jastrow--pair wavefunction involving the
molecular degrees of freedom, times a term coupling molecular and shadow positions:
\begin{equation}
K(R,S) = \prod_{i<j}\exp\left[-\frac{1}{2}\left(\frac{b}{r_{ij}}\right)^5\right]
\prod_{i=1}^N \exp\left[-C\left({\bf r}_i-{\bf s}_i\right)^2\right]
\end{equation}
The shadow degrees of freedom are also correlated by $\phi(s)$, a Jastrow product of the
same form of the one used for the molecules. The correlation  pseudopotential
is the rescaled intermolecular interaction\cite{McF94}:
\begin{equation}
\phi(s)=\exp\left[-\delta \sum_{i<j}v(\alpha s_{ij})\right].
\end{equation}
The parameters $b,C,\delta,\alpha$ appearing in the SWF are determined variationally, 
by minimizing the variance of the expectation value of the Hamiltonian 
\begin{equation}
 \langle \Delta E^2)  = \frac{\langle \Psi(R)| (H - E_0)^2\Psi(R) \rangle}
{\langle \Psi(R)|\Psi(R) \rangle}.
\end{equation}
As already mentioned, SWF provides a stable crystalline phase despite
its translationally invariant form, due to the implicitly induced
many--body correlations. The stable phase, ordered or disordered,  is determined by the variational
parameters minimizing the variance of the energy. 

For comparison we also performed VMC calculations with a standard Bijl--Jastrow--Nosanow wavefunction\cite{Nos64},
in which the atoms are constrained to remain around given lattice positions:
\begin{equation}
\Psi(R)=\exp\left[-\frac{1}{2}\sum_{i<j}\left(\frac{b}{r_{ij}}\right)^5\right]
\prod_{i=1}^{N}\exp[-C({\bf r}_i-{\bf \bar R}_i)^2],
\end{equation}
where the $\bf \bar R$ are vectors of a lattice. This wave function has been also
taken as importance function for the Diffusion Monte Carlo result that we computed
at equilibrium density.

\subsection{Simulations}

All simulations have been performed for a cell of constant volume $V$
containing a number N of molecules such that  $\rho=N/V$, and imposing 
periodic boundary conditions in order to reduce finite--size effects. 
The simulations have been performed for two different crystal geometries:
face-centered cubic (fcc) and hexagonal close-packed (hcp).
Most of the results  for the fcc crystal were obtained 
using N=108 molecules arranged on 3x3x3 elementary cubic cells of side
$a_{cell}=(4/\rho)^{\frac{1}{3}})$. For simulations in the
hcp phase we  considered N=180 molecules in a cell made up of 5x3x3 elementary 
cells. In order to check the magnitude of finite size effects 
(in particular for the vacancy formation energy) 
simulations in the fcc crystal with N=256 (4x4x4 elementary cubic cells) 
and N=500 (5x5x5 elementary cubic cells) have also been performed.

The molecule-molecule interaction is truncated at the edge of a sphere of
radius equal to half of the side of the simulation box $L$. The contribution from
the potential energy outside the sphere is estimated by assuming that the
pair correlation function is constant for distances larger than $L/2$.

The expectation values of the observables of interest have been 
computed by means of the Variational Monte
Carlo method. When using SWF, the expectation value of a local operator $\hat O(R)$
is given by:
\begin{equation}
\langle \hat O \rangle =\frac{1}{\cal N}\int\int\int\;\pi(R,S,S')\hat O(R) dR\;dS\; dS' , 
\end{equation}
where $\cal N$ is the normalization of the wave function. 
This integral can be computed sampling the joint probability distribution for 
the molecular and shadow degrees of freedom:
\begin{equation}
\pi(R,S,S') =K(R,S)\phi(S)K(R,S')\phi(S').
\end{equation}
As we illustrated in previous work\cite{Ped97}, the use
of  plain Metropolis sampling considerably slows down the convergence of the results, especially 
when the crystal includes a vacancy. This is due to the fact that the probability for a particle
to have a given displacement is conditional on the position of the corresponding
shadow degrees of freedom, to which it is connected by harmonic-like terms. 
The structure of the system is therefore analogous to that
of a collection of trimers, whose single components can have only limited relative moves.
By using pseudoforces (i.e. gradients of the function to be 
sampled) in sampling the probability distribution $\pi$, it is possible
to move all the components of a trimer together. This was demonstrated to be a very
powerful tool in the study of vacancies in solid $^4$He, where it was shown that
both relaxation of the density around the empty site and a mobility of the vacancy
can be achieved\cite{Ped97,Cha99}. The use of gradients also improves the convergence in the disordered
phase, and in metastable crystal phases, where the low density might give rise to
local disordering of the system. 
The optimization of the variational parameters appearing in the SWF was performed  
using the optimization program by C.J. Umrigar and P. Nightingale implementing a modified
Levemberg-Marquardt algorithm. 
The optimization of the parameters of the shadow--shadow correlation pseudopotential of Eq. (6)
is particularly demanding from the computational point of view. In fact the local energy does not
depend, configuration by configuration, on the value of $\delta$ and $\alpha$, but  on the
statistical weight  of the configuration $\pi(R,S,S')$ only.   
The global computational cost of our SWF, including the optimization stage, is comparable with
that of a DMC simulation for the same system. However, once a satisfactory parameterization
has been obtained, successive simulations have a cost which is at least one order of magnitude
less than DMC. For this reason SWF can be used for simulations including up to several thousands
particles, as it was done for the study of the solid-liquid coexistence in $^4$He. 

\section{Results}
\subsection{Parameterization of SWF}
The optimized variational parameters in the SWF each have a different density
dependence.  The density dependence is exploited in the Local-Density dependent version of
the SWF (LD-SWF)\cite{Ped94} that is used for describing inhomogeneous systems, like
the solid-vapor interface, with a single wavefunction.     
For such reason the optimization has to be carried out accurately in the region close to the
equilibrium density (which is experimentally found at $\rho_0=0.02595$\AA$^{-3}$\cite{vSt88}), and lower. 
The parameters $b$, connected to  the width of the correlation hole of the molecules, and $C$,
giving the inverse of the mean square displacement between particles and shadows, show a weak dependence
on the density. In particular the parameter $C$ is nearly constant and equal to 1\AA$^{-2}$
throughout the density range considered. The parameter $b$ tends to saturate at high densities around
a value $3.35$\AA, while at equilibrium density its value is $b=3.25$\AA. For densities lower
than the equilibrium density the parameter $b$ shows some oscillations, contained within 5\% of the 
value at equilibrium value. Such oscillations might be due to the fact that the system  and
autocorrelations between configurations become very large, making the optimization difficult.  
For such reasons  a reasonable choice in the LD-SWF 
is to keep the parameters $b$ and $C$ constant at the value optimal at the equilibrium density. 
The remaining variational parameters $\alpha$ and $\delta$ show a stronger density dependence. 
As already mentioned the energy at a given density has a much weaker dependence on the
values of the shadow--shadow correlation parameters, due to the fact that
$\phi(S)$ does not enter in the estimate of the local energy. 
In Fig. 1 we show the results of the optimization. The
errorbars indicate the range of values for which the variation in the computed variational
energy is within two standard deviations from the optimal value. 
\begin{figure}

\includegraphics[scale=0.5]{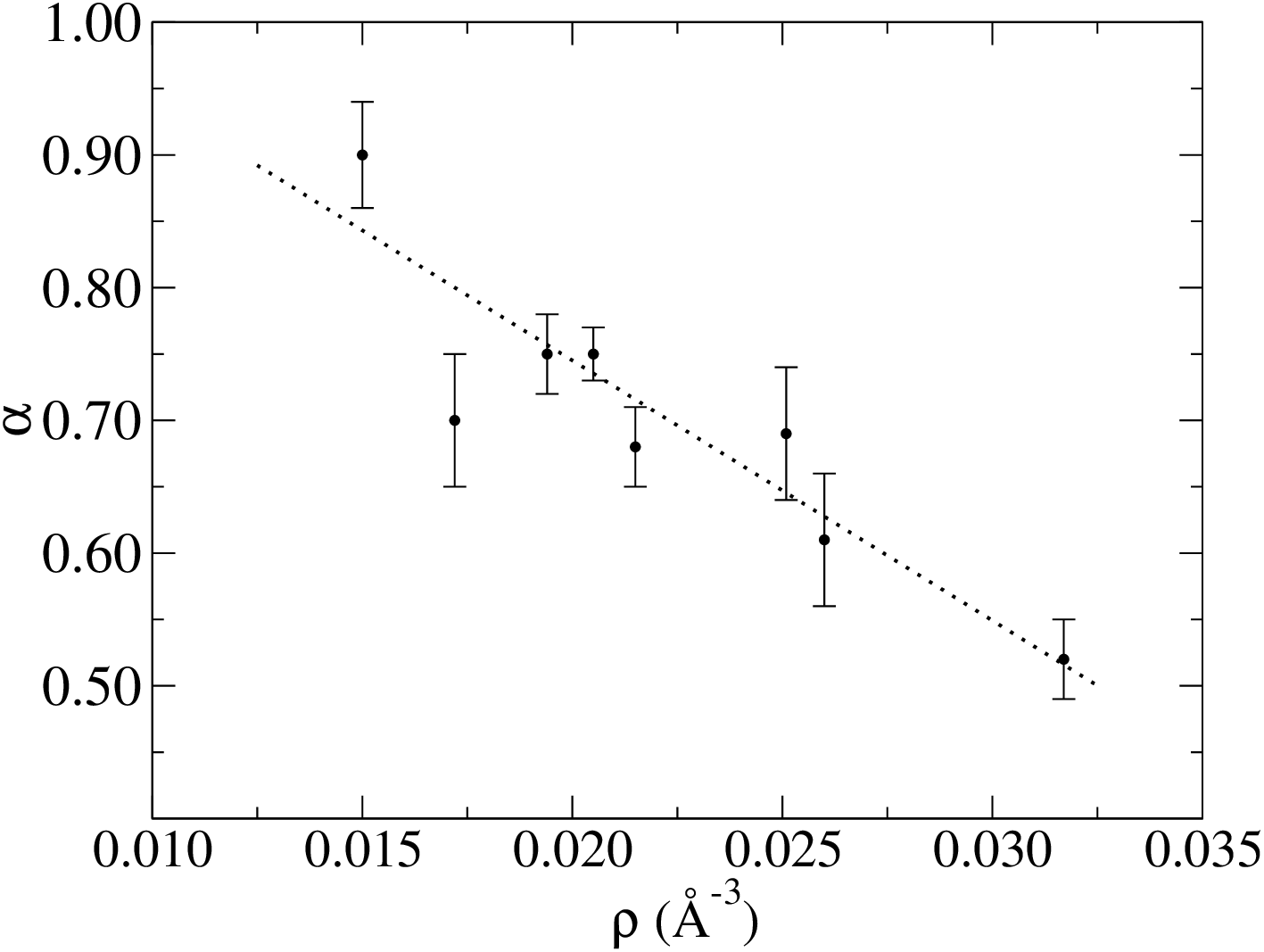}

\vspace{2cm}
\includegraphics[scale=0.5]{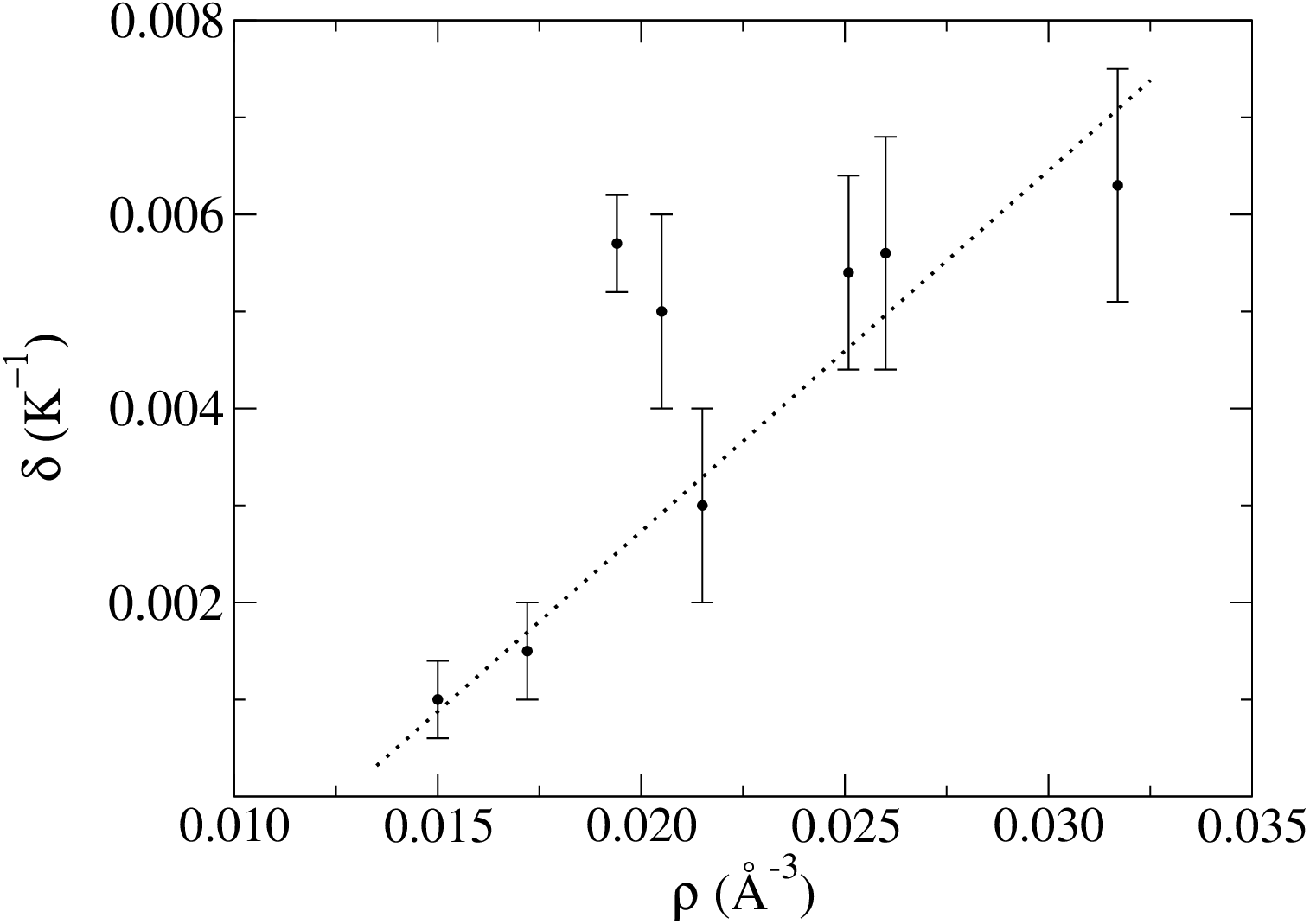}
\caption{Dependence of the SWF variational parameters $\alpha$ (upper panel) and $\delta$
(lower panel)  on the molecular density. Lines show the linear fits to the computed values.}
\end{figure} 
\begin{table}
\begin{center}
\begin{tabular}{cc}
\toprule
$\alpha_0$&1.2049$\pm$0.0872\\
$\alpha_1$&-21.961$\pm$2.984\\
$\delta_0$&-0.0047$\pm$0.0008\\
$\delta_1$&0.3717$\pm$0.0370\\
\botrule
\end{tabular}
\end{center}
\caption{Values of the coefficients of Eq. (11) for the fits of the shadow-shadow
pseudopotential parameters $\alpha$ and $\delta$ as a function of the density.
}
\end{table}
These parameters need to be made functions of the density in  the LD-SWF. 
It is reasonable to  assume 
a linear dependence of the parameters $\alpha$ and $\delta$ on the density 
\begin{eqnarray}
\alpha(\rho)&=&\alpha_0+\alpha_1\rho \nonumber \\
\delta(\rho)&=&\delta_0+\delta_1\rho. 
\end{eqnarray}
The coefficients of the linear fits  are given  in Table II. The interpolations are also plotted in Fig. 1.
The differences between energies computed with the optimal values of the parameters  and 
the values for the LD-SWF parameterization less than 1\%.

\subsection{Ground state properties}
The SWF has been used to compute the  
variational energies for molecular \pH. The results are reported in tables III-IV.
In table III we compare the binding energy for the density $\rho_0=0.02609$\AA$^-3$,
(corresponding to a specific volume $v=23.08$cm$^3\;$mol$^{-1}$)
obtained by SWF, variational Monte Carlo with a Jastrow-Nosanow wave function (JN)
and Diffusion Monte Carlo (DMC). We also compare to the experimental value reported in Ref. \onlinecite{Nor84}
As it can be seen the energy obtained by SWF improve the Jastrow--Nosanow result.
Nevertheless the difference between the SWF and DMC results is still larger than 1K. 
The DMC result at this density shows a quite large discrepancy (about 6\%) with respect
to the available experimental results. This is in agreement with the findings of 
Norman et al., who point out that other model interactions give a better estimate 
of the binding energy near the equilibrium density. 
\begin{table}
\begin{center}
\begin{tabular}{cc}
\toprule
&E/N(K)\\
\colrule
SWF&-86.679(3)\\
JN&-86.214(2)\\
\colrule
DMC&-87.931(5)\\
Exp.&-93.5\\
\botrule
\end{tabular}
\end{center}
\caption{Ground state energies per particle at the density $\rho_0=0.02609$\AA$^{-3}$
obtained by VMC with SWF and JN, compared to the DMC result and the experimental
value from Ref.\onlinecite{Nor84}}  
\end{table}

\begin{table}
\caption{Total, potential, and kinetic energies per particle, in K,  as a function of density from  
VMC-SWF calculations. For densities at which the system was found to be a crystal, the
values are listed for the fcc and hcp lattices (obtained with N=108 and N=180 molecules respectively).}
\squeezetable
\begin{tabular}{ccccccc}
\toprule
$\rho$& $E_0/N$ &$T/N$&$V/N$&&&\\
\colrule
0.0050& -17.400(16) &4.139(15)&-21.540(10)&&&\\
0.0080& -29.348(13)&8.202(16)&-37.476(10)&&&\\
0.0110& -41.390(9)&14.269(14)&-55.543(7)&&&\\
0.0150& -55.929(5)&24.037(5)&-79.966(3)&&&\\
0.0172& -63.800(7)&34.019(9)&-97.713(6)&&&\\
\colrule
&&fcc&&&hcp&\\
&$E_0/N$&$T/N$&$V/N$&$E_0/N$&$T/N$&$V/N$\\
\colrule
0.01940& -70.862(5)&43.469(5)&-114.332(7)&-71.288(6)&43.499(11)&-114.787(11)\\
0.02150& -79.553(2)&52.029(3)&-131.582(3)&-79.597(6)&52.665(9)&-132.262(7)\\
0.02509& -86.354(2)&66.541(4)&-152.895(4)&-86.494(4)&66.502(9)&-152.996(8)\\
0.02600& -86.679(3)&71.400(5)&-158.079(5)&-86.813(5)&71.395(15)&-158.209(15)\\
0.02900& -83.425(4)&83.868(9)&-167.293(11)&-83.571(7)&83.877(19)&-167.448(21)\\
0.03170& -73.296(5)&97.142(10)&-170.438(12)&-73.442(9)&97.111(19)&-170.553(23)\\
0.03400& -58.691(8)&109.237(18)&-167.927(18)&-58.484(3)&109.375(5)&-167.858(5)\\
0.03700& -30.281(11)&122.924(22)&-153.206(25)&-30.139(2)&122.976(4)&-153.115(5)\\
0.04000& 8.861(2)&141.570(3)&-132.708(4)&8.931(2)&141.450(5)&-132.519(2)\\
\botrule
\end{tabular}
\end{table}

In Table IV we report the variational results obtained with our SWF in the disordered phase and 
in the crystalline phase, for which values for the fcc and hcp crystals are given.
The results are also plotted in Fig. 2.
\begin{figure}
\includegraphics[scale=0.6]{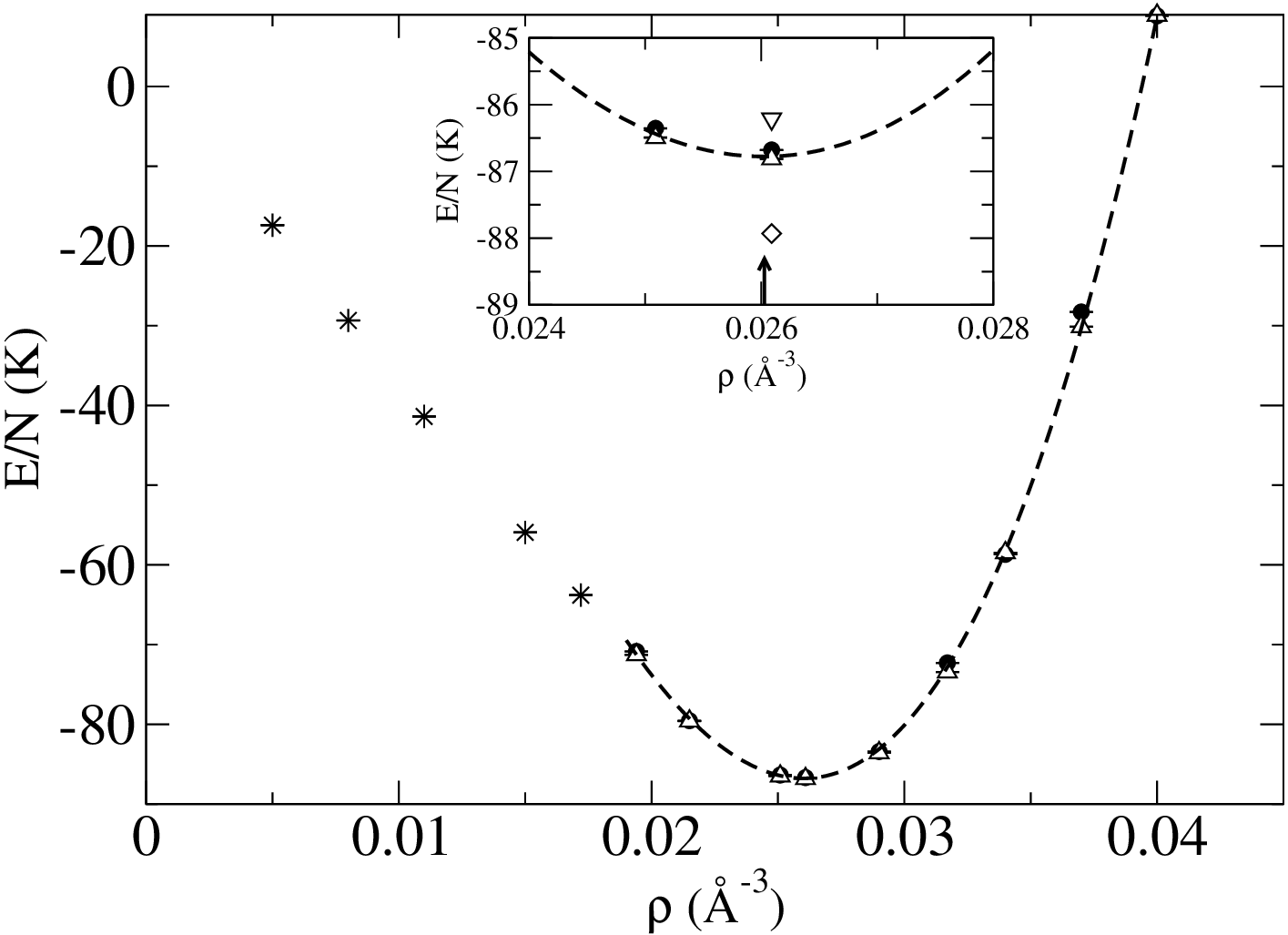}
\caption{Estimated energy per molecule in solid and liquid \pH. Filled circles: SWF results for
the fcc crystal; empty triangles: SWF results for the hcp crystal; stars: SWF results in the 
disordered phase. The inset gives an expanded view around the equilibrium density. Filled triangle: JN
result; diamond: DMC result. The dashed line is a plot of the fit with Eq. (13). The arrow indicates
the equilibrium density obtained by the fit.}  
\end{figure}
The binding energy for the hcp lattice is lower than for the fcc
lattice for densities $\rho<0.03170$\AA$^{-3}$.   

As already mentioned, when using  shadow wavefunctions,
the stable phase (crystalline or disordered) is determined entirely from  the variational principle.
Therefore, given a set of variational parameters, we do not know {\it a priori} if the molecules will
stay localized on a lattice. In order to determine the phase of the system, we need to compute
a crystalline order parameter of this form:
\begin{equation}
O=\frac{1}{NM}\sum_{i=1}^{N}\sum_{\alpha=1}^{M}\left|e^{-i\bf k_\alpha\cdot\bf r_i}\right|,
\end{equation}
where the ${\bf k}_\alpha, \alpha=1...M$ are vectors of the reciprocal lattice for which the order
is monitored. This quantity
is exactly 1 if all the molecules are sitting on top of a lattice site, while in a disordered
system its value is about 1/$\sqrt{N}$, where $N$ is the number of particles. 
The value we find at the equilibrium density is  $\langle O\rangle\sim 0.72$ for the molecular degrees of freedom. 
This value slightly increases with the density of the system as the molecules become more and more localized. 
At the highest density considered we found $\langle O \rangle \sim 0.76$.
The limit  of stability for the crystal phase is found at a density $0.0172$\AA$^{-3} < \rho  < 0.0194$\AA$^{-3}$.  

In order to compute the  value of the equilibrium density and pressure, it is
convenient to fit the results for the ground state energy per particle by means of the following 
expression:
\begin{equation}
E(\rho)=E_0+a\rho+b\rho^\gamma .
\end{equation} 
The coefficients for the solid phase in the fcc and hcp crystals 
are reported in table V. The estimate of the
\begin{table}
\begin{tabular}{cccccc}
\toprule
&$E_0$&$a$&$b$&$\gamma$&$\rho_0$ (\AA$^{-3}$)\\
\colrule
fcc&78.8502&-9205.32&1.02128$\times 10^7$&3.24389&0.02601\\
hcp&82.8656&-9478.60&9.41995$\times$10$^6$&3.21149&0.02603\\
\botrule
\end{tabular}
\caption{Coefficients of the equation of state (Eq. (13)) fitted to SWF
simulation data. The energy is given in K; $\rho_0$ is the
computed equilibrium density.} 
\end{table}
equilibrium density is in good agreement with the experimental finding $\rho_0^{exp}=0.02595$\AA$^{-3}$\cite{vSt88}.
The pressure as a function of density is then obtained from the following expression:
\begin{equation}
P(\rho)=\frac{1}{\rho^2}\frac{\partial E(\rho)}{\partial \rho}.
\end{equation}
\begin{figure}
\includegraphics[scale=0.6]{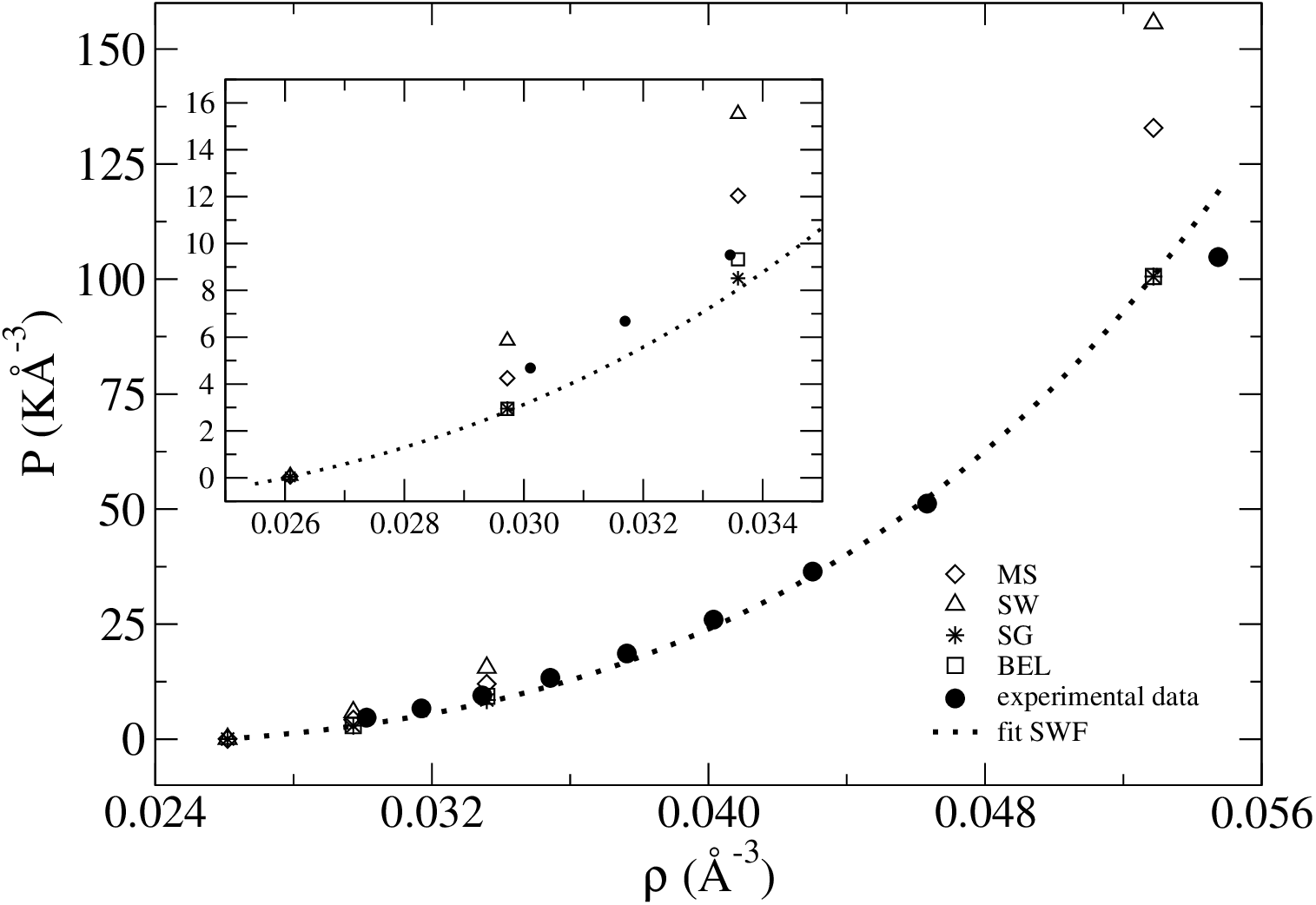}
\caption{Pressure as a function of density in solid \pH. 
Dotted line: SWF result computed from Eq. (15)
from the fit of the ground state energy. Filled circles are experimental data
from ref. \cite{vSt88}. Points are values computed in ref. \cite{Nor84} with different
intermolecular potentials (MS: Meyer and Schaefer\cite{Sch79}; SW: Schaefer and Watts\cite{Sch82};
SG: Silvera and Goldman\cite{Sil78}; BEL: Buck {\it et al.}\cite{Buc83}). In the inset
we expand the same data near the equilibrium density.        
}
\end{figure}
In Fig. 3 the computed pressure  for the fcc and hcp solid phases is compared to the
experimental curve extrapolated to T=0\cite{vSt88}. The SWF result is lower than the experiment
near the equilibrium density, while the pressure tends to be overestimated at high 
densities. We also compare our results with results of
Norman et al.\cite{Nor84} for the same Silvera-Goldman potential and for 
other intermolecular interactions.   

\begin{figure}
\includegraphics[scale=0.25]{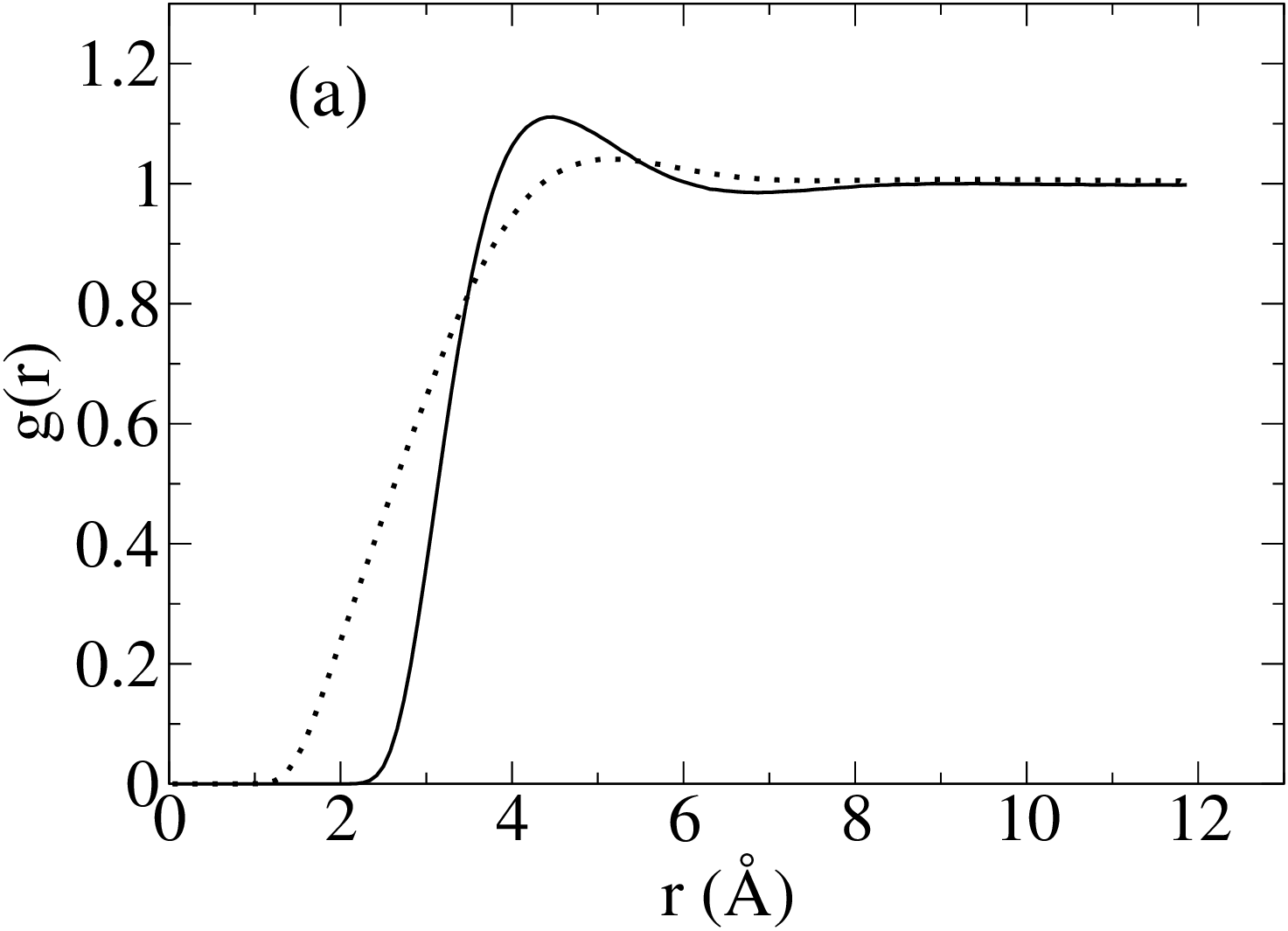}\hspace{0.5cm}
\includegraphics[scale=0.25]{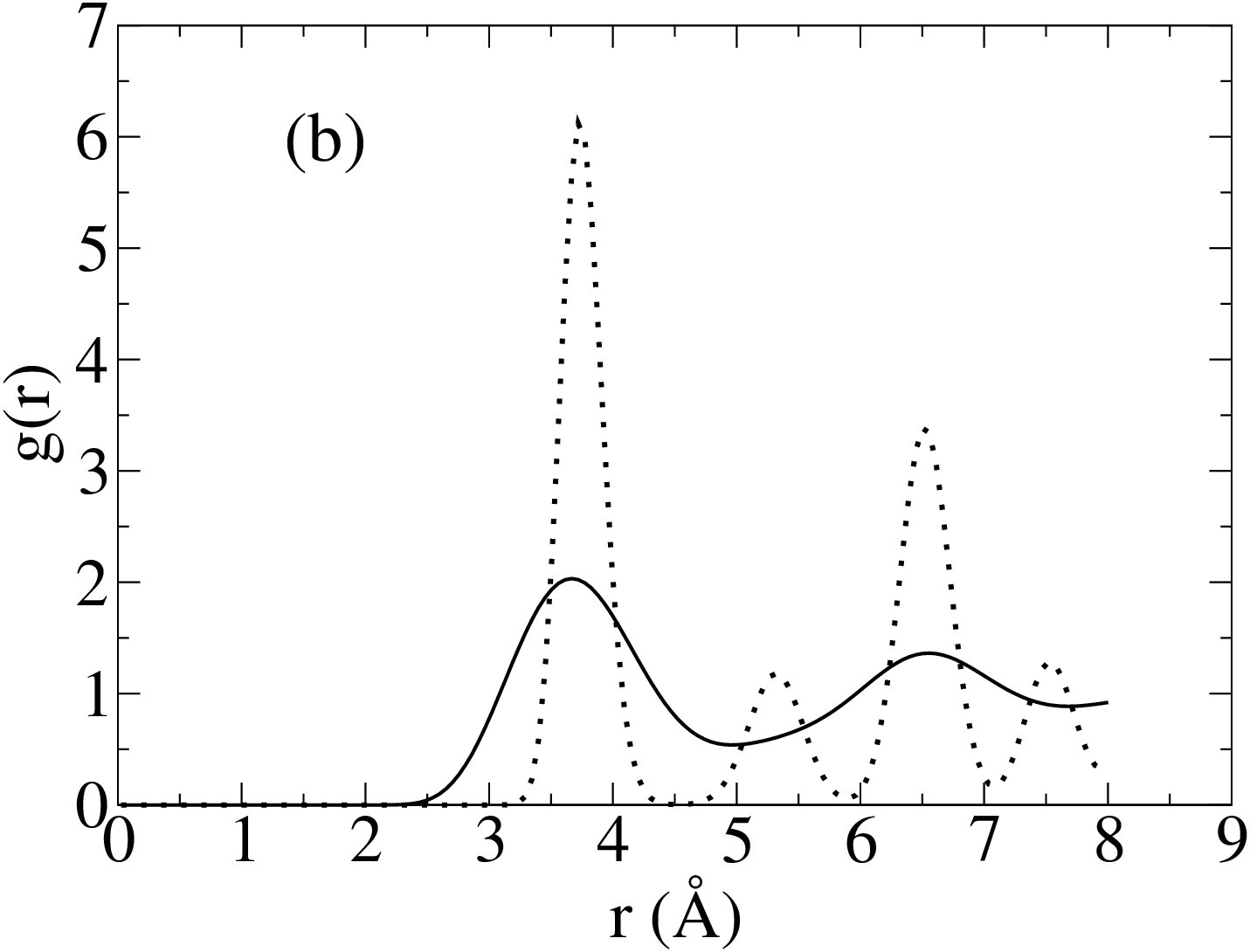}

\vspace{0.7cm}
\includegraphics[scale=0.25]{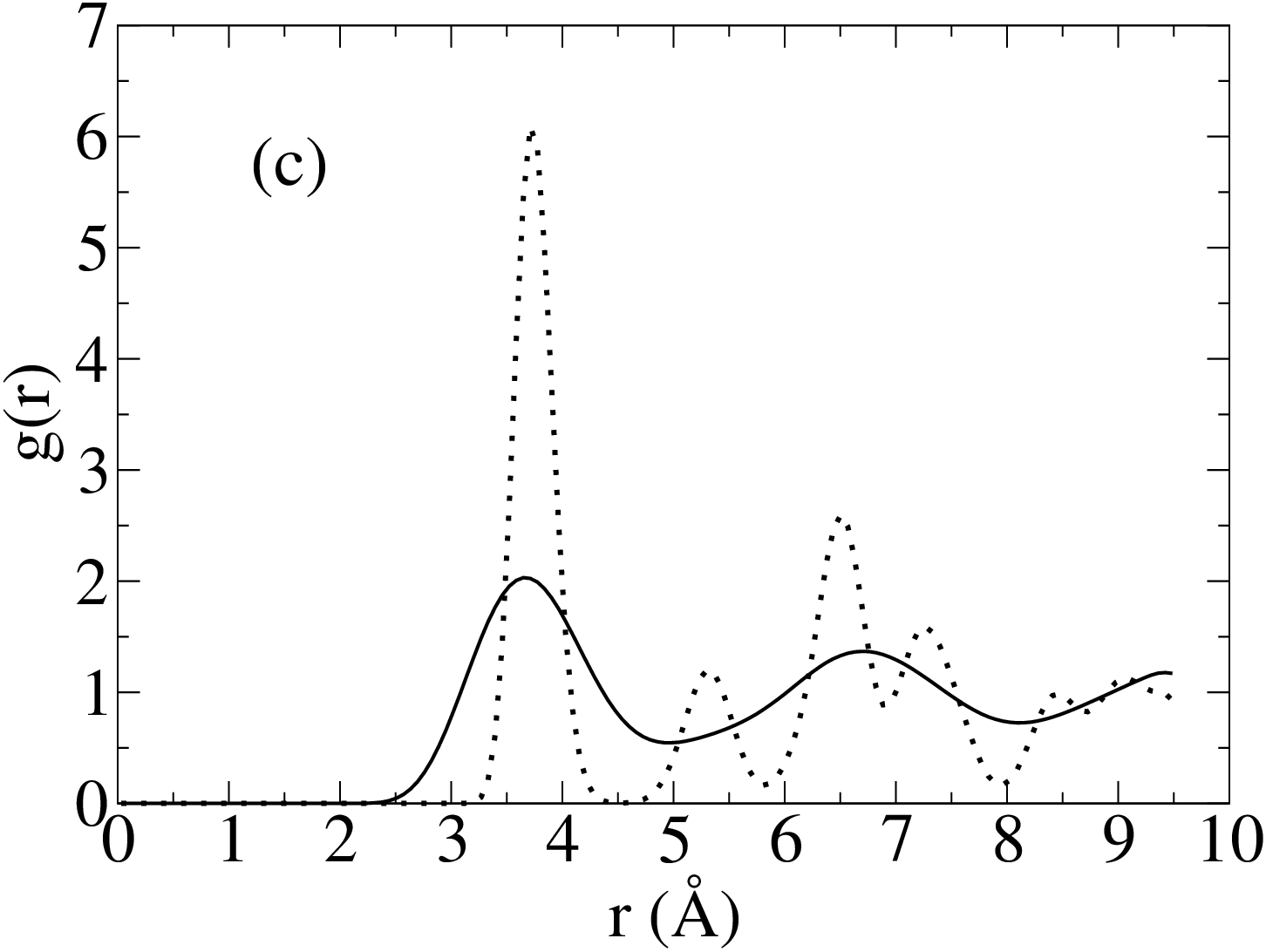}\hspace{0.5cm}
\includegraphics[scale=0.25]{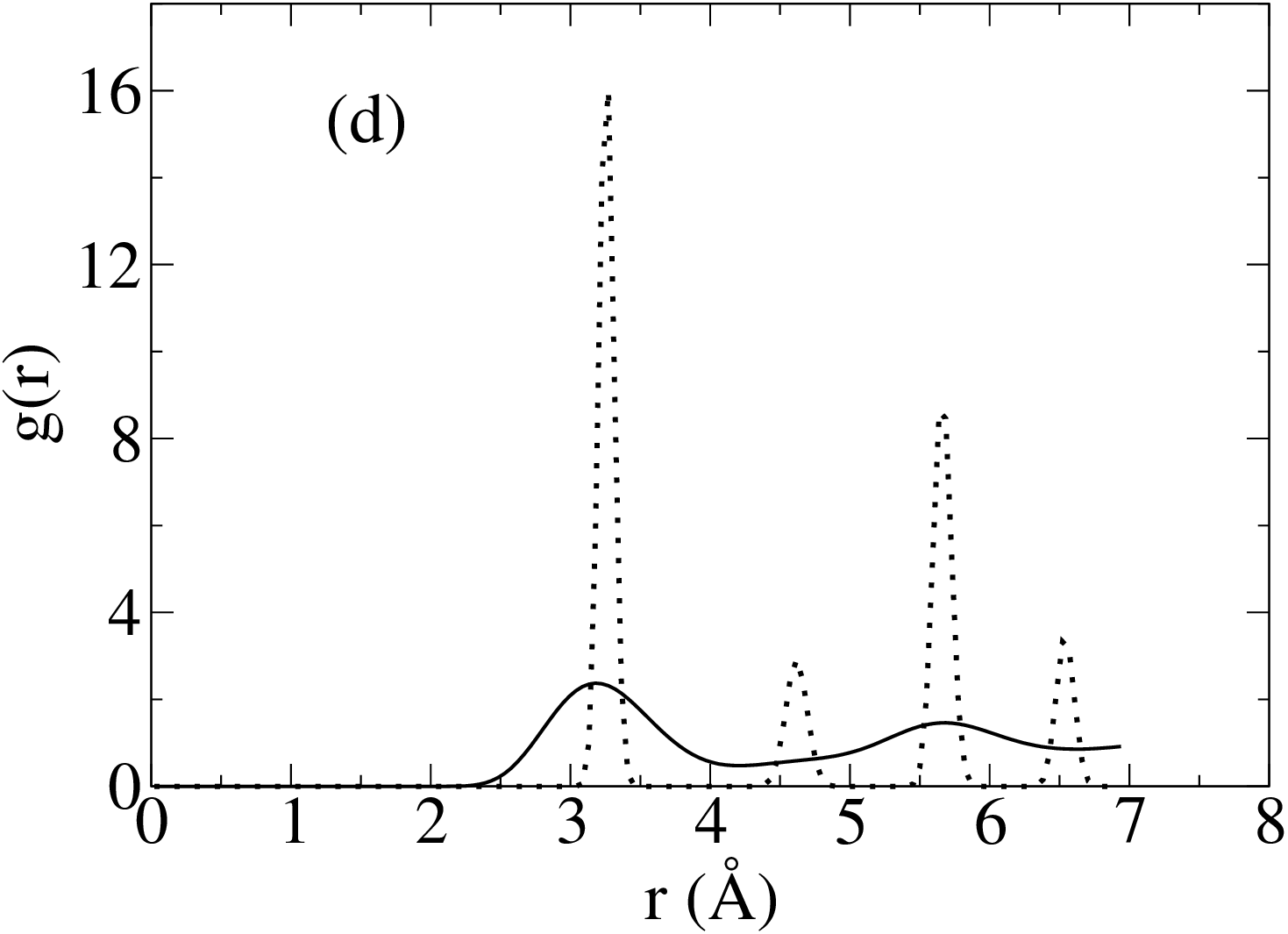}
\caption{Pair distribution functions computed with SWF in molecular \pH.
Solid lines: pdf of particles; dotted line: pdf for shadows.
(a) $\rho=0.00800$\AA$^{-3}$ (disordered phase); 
(b) $\rho=0.02609$\AA$^{-3}$ (fcc crystal); 
(c) $\rho=0.02609$\AA$^{-3}$ (hcp crystal); 
(d) $\rho=0.04000$\AA$^{-3}$ (fcc crystal). 
}
\end{figure}
In Fig. 4 we plot the pair distribution function:
\begin{equation}
g(r)=\langle \sum_{i\neq j}\delta(|{\bf r}_i-{\bf r}_j|,r)\rangle,
\end{equation}
were $\langle \rangle$ indicates the expectation value over the SWF.
The same quantity can be defined for the shadow degrees of freedom: 
\begin{equation}
g_s(r)=\langle \sum_{i\neq j}\delta(|{\bf s}_i-{\bf s}_j|,r)\rangle.
\end{equation}
$g(r)$ and $g_s(r)$ are displayed  for a density below $\rho_0$, at $\rho_0$ for the
fcc and hcp crystals and at an higher density for the fcc crystal. At the lowest density 
$g(r)$ has the typical behavior that can be observed in a fluid, with a peak at the
first shell of about 1.1. At larger densities, where the system is crystallized, 
it is possible to observe how the first peak increases to values around 2. The peaks
corresponding to the second and third shell are nearly merging into each other. 
This feature,  common to other quantum solids, is 
due to the wide dispersion of the molecules around the lattice sites. 
On the other hand, the distribution of shadows presents, as expected,  a much better defined structure. 
The peaks typical of the fcc and hcp structures, already well visible at
density $\rho_0$ become separated at higher densities. 
The shadow--shadow correlation induces a 
much stronger localization, and  at higher densities the distribution becomes 
similar to that of a classical solid.  

\subsection{Vacancies}

The vacancy formation energy at constant pressure in a system of N particles and $N_l$ lattice sites
at a given density $\rho$ can be defined as\cite{Guy72,Ped297}:
\begin{equation}
\Delta E_{vac}=E(V',N=N-1,N_l=N)-\frac{N-1}{N}E(V,N,N_l=N).
\end{equation}
The volumes $V$ and $V'$ are related 
by the requirement that the density of the system remains constant,
i.e., $N/V=(N-1)/V'$. In a quantum crystal $\Delta E_{vac}$ has contributions from three different
effects. The main contribution comes from  the missing potential and kinetic energy due to the 
missing particle. At constant pressure and in the thermodynamic limit, this
energy is a function of the pressure and of the potential energy per particle:  
\begin{equation}
\Delta E_{vac}^{0} = \frac{1}{\rho}P(\rho)-\frac{V(\rho)}{N}.  
\end{equation}
Another important contribution comes from the relaxation of the crystal around the
empty site. This term lowers the vacancy formation energy due to the fact that
the kinetic energy of the molecules surrounding the vacancy is reduced. This contribution
is more important at lower densities, and can be well described only within a model that
allows lattice sites to be displaced from their original positions, as  happens for SWF where
no {\it a-priori} lattice is assumed. 
The third contribution comes from 
the motion of the vacancy through the crystal. The bandwidth of the vacancy has recently been recently 
estimated  to be 6 to 10K in $^4$He by Galli and Reatto\cite{Gal03}.
We expect this contribution to be smaller in \pH, due to  the higher degree of localization 
of the molecules.
 
In Fig. 5 we report the results obtained for $\Delta E_{vac}$ as a function of the
density for the fcc crystal (obtained with N=108 molecules), and for the hcp crystal
(obtained with N=180 molecules).
The values are compared with the corresponding estimate of $\Delta E_{vac}^{0}$. 
\begin{figure}
\includegraphics[scale=0.6]{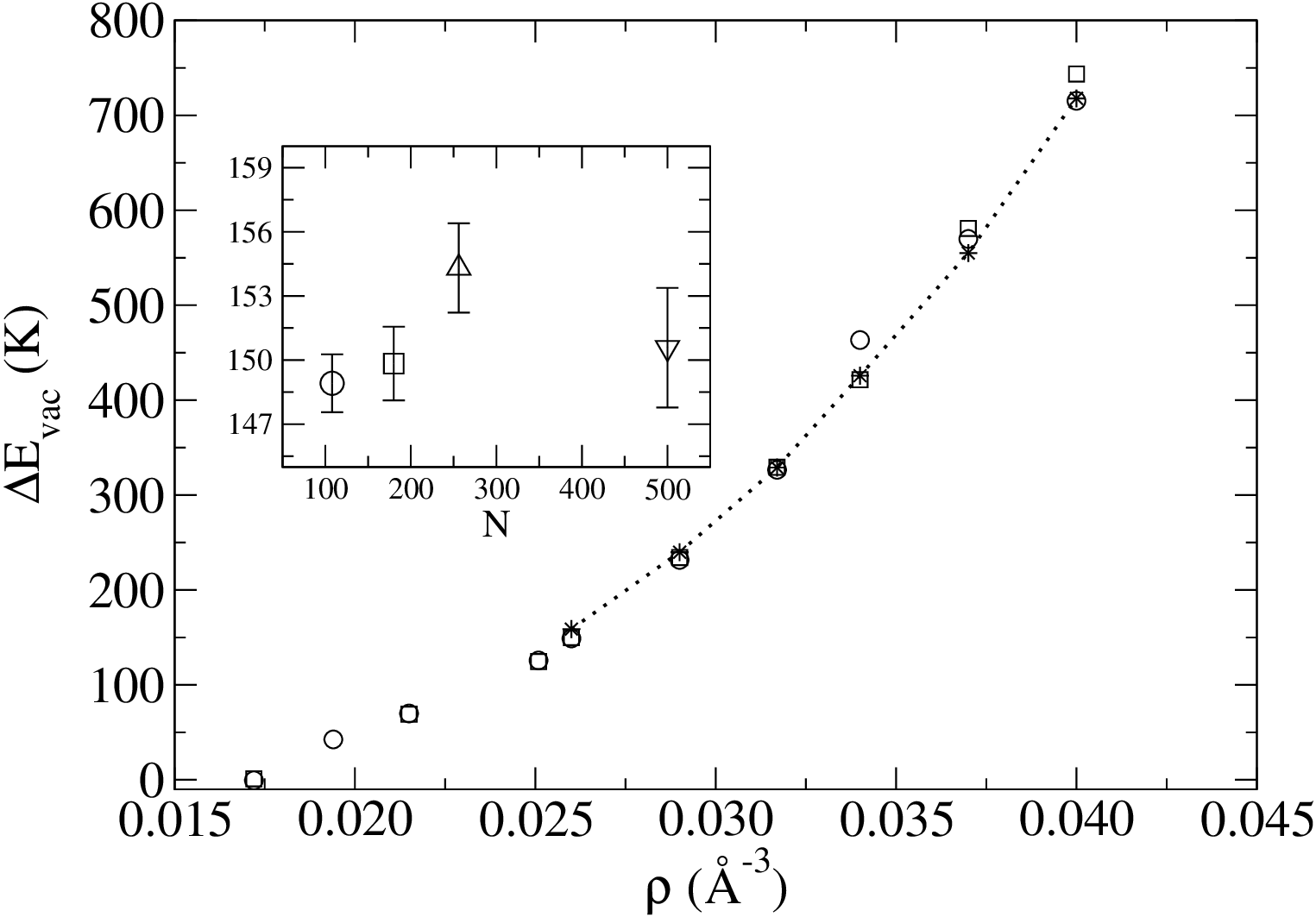}
\caption{Vacancy formation energy $\Delta E_{vac}$ from SWF calculations as a function
of the molecular density. Circles: fcc crystal (108/107 particles);
squares: hcp crystal (180/179 particles). The dotted line represents the 
estimated energy for a vacancy without relaxation effects, obtained from the 
fitted equation of state.
In the inset we plot the vacancy formation energy at density 
$\rho=0.02609$\AA$^{-3}$ as a function of $N$, the number of particles used in the
simulation.}
\end{figure}
As  can be seen there is
a strong dependence of $\Delta E_{vac}$ on the density. For low densities the contribution
from lattice relaxation, which can be estimated by the difference $E_{rel}=\Delta E^{vac}_0 -\Delta E^{vac}$,
is $E_{rel}\sim 9K$. This value decreases when the density increases. For
high densities $E_{rel}$ does not show a systematic behavior. The estimate of the
energy of a static vacancy depends on 
the potential energy
per particle, which has a larger uncertainty at higher densities due to 
larger autocorrelations and stronger finite-size effects.
Finite size effects also  make the estimate of $\Delta E_{vac}$ from the variational results 
harder to obtain. This can also be inferred from the fact that the results for the fcc and hcp lattices tend to depart
from each other in a non systematic way. 
Their difference can be considered as a measure of the actual uncertainty on the
computed value of $\Delta E_{vac}$
The use of periodic boundary conditions implies that  the presence of one vacancy in the
simulation cell corresponds to a density 1/N of vacancies in the infinite crystal. 
The local deformation of the crystal due to the relaxation of the atoms around the empty site
might induce deformations to the next shells, which  might translate into a vacancy-vacancy
interaction. In order to check if such interaction affects the estimate of the 
vacancy formation energy, we performed simulations at different numbers of molecules at the
equilibrium density (where relaxation effects are large),
and recomputed the vacancy formation energy. Results are reported in table VI. As 
can be seen, the differences are within two standard deviations,
suggesting that no sizeable effects of the
relaxation can be observed beyond the second shell of neighbors (which is contained in
the simulation box with 108 molecules).   
This is also confirmed by the fact that at density $\rho_0$ no significant difference is observed for the vacancy formation
energy in the fcc and the hcp crystals, that have the same structure for the first
shell of neighbors. 
\begin{table}
\begin{center}
\begin{tabular}{cc}
\toprule
N&$\Delta E_{vac}$ (K)\\
\colrule
107& 149(1)\\
179& 150(2)\\
255& 154(2)\\
499& 150(3)\\
\botrule
\end{tabular}
\end{center}
\caption{Vacancy formation energy for \pH at density
$\rho=0.02600$\AA$^{-3}$ as a function of the number of particles used in the simulation.}
\end{table}

\section{Conclusions}

A shadow wavefunction has been devised to study ground state properties of molecular
para-hydrogen, using the Silvera--Goldman interaction. The equation of state
of the crystalline fcc and hcp phases has been computed optimizing the variational
parameters of the SWF, and fitting them as functions of the
density. The estimated energy per particle is on average 10\% higher than the available
experimental data. However the estimate of the pressure and of the equilibrium density
suggest that this SWF gives a realistic description of the system, allowing for future
developments, in particular for as regards the study of the solid--vapor interface, and
the study of clusters. 
The vacancy formation energy has been estimated in a wide range of densities. At densities
around the equilibrium value relaxation effects appear to be about 5\% of the total
vacancy formation energy.


\begin{acknowledgments}
We would like to thank G.V. Chester, M.H. Kalos, S. Fantoni, and L. Reatto  for useful comments  
and suggestions about this work. We also thank C.J. Umrigar for providing us the
Levemberg-Marquardt package used for the optimization of the parameters in the
Shadow Wave Function. Part of this work was supported by  a grant COFIN2002 
"Quantum Fluids and Solids in Confined Geometries" of the 
Italian Ministero dell' Istruzione, Universit\`a e Ricerca. Calculations have been performed 
in part on the Beowulf cluster in CINECA under a INFM parallel computing grant, 
and in part on the ALPS cluster at ECT* in Trento.
\end{acknowledgments}


%
%

%
%

\end{document}